\documentclass[%
 reprint,
 amsmath,amssymb,
 aps,
]{revtex4-2}

\usepackage[utf8]{inputenc}

\usepackage[T1]{fontenc}
\usepackage{graphicx}
\usepackage{dcolumn}
\usepackage{bm}
\usepackage{siunitx}
\usepackage{chemformula}
\usepackage{txfonts}
\usepackage{amsmath}
\usepackage{float}
\usepackage{hyperref}

\begin{document}

\preprint{APS/123-QED}


\title{High fidelity flopping-mode single spin operation \\
with tuning inter-dot orbital levels}

\author{Y. Matsumoto$^{1}$}
\email{y.matsumoto@tudelft.nl}
\author{X.-F. Liu$^{1}$}
\author{A. Ludwig$^{4}$}
\author{A. D. Wieck$^{4}$}
\author{K. Koike$^{7}$}
\author{T. Miyoshi$^{5,7}$}
\author{T. Fujita$^{1,2,3,5,6}$}
\author{A. Oiwa$^{1,2,3,5,6}$}

\affiliation{$^{1}$SANKEN, The University of Osaka, 8-1 Mihogaoka, Ibaraki, Osaka 567-0047, Japan \\ $^{2}$Center for Spintronics Research Network, Graduate School of Engineering Science, The University of Osaka, 1-3 Machikaneyama, Toyonaka, Osaka 560-8531, Japan \\ $^{3}$Artificial Intelligence Research Center, SANKEN, The University of Osaka, 8-1 Mihogaoka, Ibaraki, Osaka 567-0047, Japan \\ $^{4}$Lehrstuhl f{\"u}r Angewandte Festk{\"o}rperphysik, Ruhr-Universit{\"a}t Bochum, Universit{\"a}tsstra{\ss}e 150, Geb{\"a}ude NB, D-44780, Germany \\ $^{5}$Center for Quantum Information and Quantum Biology, The University of Osaka, 1-2 Machikaneyama, Tokyonaka, Osaka 560-0043, Japan\\$^{6}$Spintronics Research Network   Division, Institute for Open and Transdisciplinary Research Initiatives, The University of Osaka, 2-1 Yamadaoka, Suita, Osaka 565-0871, Japan$^{7}$e-trees.Japan, Inc. 2-9-2 Owadacho Hachioji 192-0045m, Japan \\} 




\date{\today}

\begin{abstract}

Fast spin manipulation and long spin coherence time in quantum dots are essential features for high fidelity semiconductor spin qubits. However, generally it has not been well established how to optimize  these two properties simultaneously, because these two properties are usually not independent from each other. Therefore, the scheme for  high fidelity operation by simultaneous tuning Rabi frequency and coherence time, which does not rely on the material-dependent strong spin-orbit interaction and the local magnetic field gradient limiting their scalability, are strongly demanded. Here, we demonstrate an approach to achieve high-fidelity spin control by tuning inter-dot spin-orbit coupling in a GaAs triple quantum dot (TQD), where the third dot provides precise control over orbital energy levels. In an electrically stable charge state with optimized tunnel coupling, we achieve Rabi frequencies exceeding 100 MHz while maintaining coherence through proper tuning of the inter-dot orbital levels of the TQD. By implementing a machine learning-based feedback control that efficiently estimates qubit frequency using past measurement data, we characterize and mitigate the impact of low frequency noise on qubit coherence with minimal measurement overhead. Finally, we demonstrate a $\pi$/2 gate fidelity of 99.7\% with a gate time of 4 ns through randomized benchmarking, even in a GaAs quantum dot device where electron spin coherence is typically limited by strong hyperfine interaction with nuclear spins. Our approach provides a scalable strategy for high-fidelity spin control in semiconductor quantum dot arrays by utilizing device-specific parameters rather than relying on material properties or external field gradients.

\end{abstract}

\maketitle


\section{\label{sec:Introduction}Introduction}

Semiconductor quantum dots (QDs) reveal a promise for scalable quantum computing owing to their long coherence time\cite{iso,Si,singlef}, high-fidelity quantum operations\cite{fide1,fide2,fide3,fide4}, potential for elevated temperature operation\cite{1K,1K2}, and compatibility with industrial manufacturing\cite{intel,qdesign,qdesign2}. While single-qubit control fidelities exceeding the fault-tolerance threshold of 99\%\cite{threshold} have been demonstrated, uniformity of the high-fidelity control across large qubit arrays remains challenging.

Standard approaches for fast electrical spin control rely on either intrinsic spin-orbit coupling\cite{Nowack,SOI,Ge} or field gradients induced by micromagnet\cite{MM,yoneda}. High fidelity control has been realized by balancing the gate speed and coherence time\cite{Si}, as faster control brings larger susceptibility to noises while longer coherence time needs less coupling to environments. This balancing makes optimizing for high-fidelity spin qubit operations not straightforward and complex, particularly as the number of qubits increases. Moreover, both approaches have fundamental scaling limitations: spin-orbit based control depends heavily on materials\cite{SOI,Ge,Ge2}, while micromagnet approaches need precise geometric positioning that becomes increasingly difficult to maintain over large arrays\cite{design,mm_opt}. These limitations highlight the need for control methods that utilize only electrically tunable parameters, allowing individual optimization for each qubit independent of the device-fixed parameters.

Alternative approaches based on hybridization of spin with another degree of freedom, valley\cite{valley} or orbital excited states\cite{ondemand} have been demonstrated for high-fideliy control, but face the challenges in optimizing the shape of the hybridization gap due to limited tunability of energy levels in single-dot devices. The flopping-mode architecture in double quantum dots (DQDs)\cite{flopping,flopping4} offers a more flexible solution, resulting in an enhanced Rabi oscillation driven by inter-dot tunneling near zero detuning. This scheme enables the control of spin-charge hybridization. However, high-fidelity control by optimizing gate speed and coherence time has not fully established yet. Specifically, factors such as the magnitude of charge noise and the magnetic field gradient mean that optimizing the tunnel coupling and detuning parameters is not always sufficient to simultaneously achieve fast gate operations and long coherence times for high-fidelity operation. Additional tuning parameters are therefore needed for
addressing this optimization bottleneck.

We propose the extended flopping-mode architecture to triple quantum dots (TQDs), where the third QD provides an additional tunable parameter: the energy levels of the third dot. We demonstrate that this enhanced tunability of the flopping-mode qubit through the coupling to the third dot enables precise engineering of the shape of the hybridization gap through the orbital energy structure of TQDs while maintaining the fundamental advantages of the flopping-mode scheme. The triple quantum dot (TQD) architecture allows post-optimization in electrical manners for each qubit, unlike approaches relying on device-fixed parameters like spin-orbit interaction or micromagnets in a single QD.

Moreover, understanding and mitigating qubit frequency fluctuations in individual qubits remains essential for characterizing and improving multi-qubit devices. In particular, the impact of noise on the effective spin splitting through inter-dot charge states in flopping-mode operation has not been well characterized. Although conventional measurement-based feedback approaches using Bayesian estimation\cite{feedback,feedback2} can trace these fluctuations, numerous measurements per qubit needed for these estimations, would make these estimation impractical for characterizing noise in multi-qubit systems where correlated noise effects become important\cite{noise,noise2}. While dynamical decoupling techniques\cite{echo,CPMG,CPMG2} can extend coherence times, they require modifications to the target operations and may not be suitable for all quantum algorithms. As quantum dot arrays scale up, efficient methods for noise characterization and control that minimize measurement overhead become increasingly crucial, both for understanding device performance and for implementing practical feedback schemes in larger systems.

\begin{figure*}
\centering
\includegraphics[width=16cm]{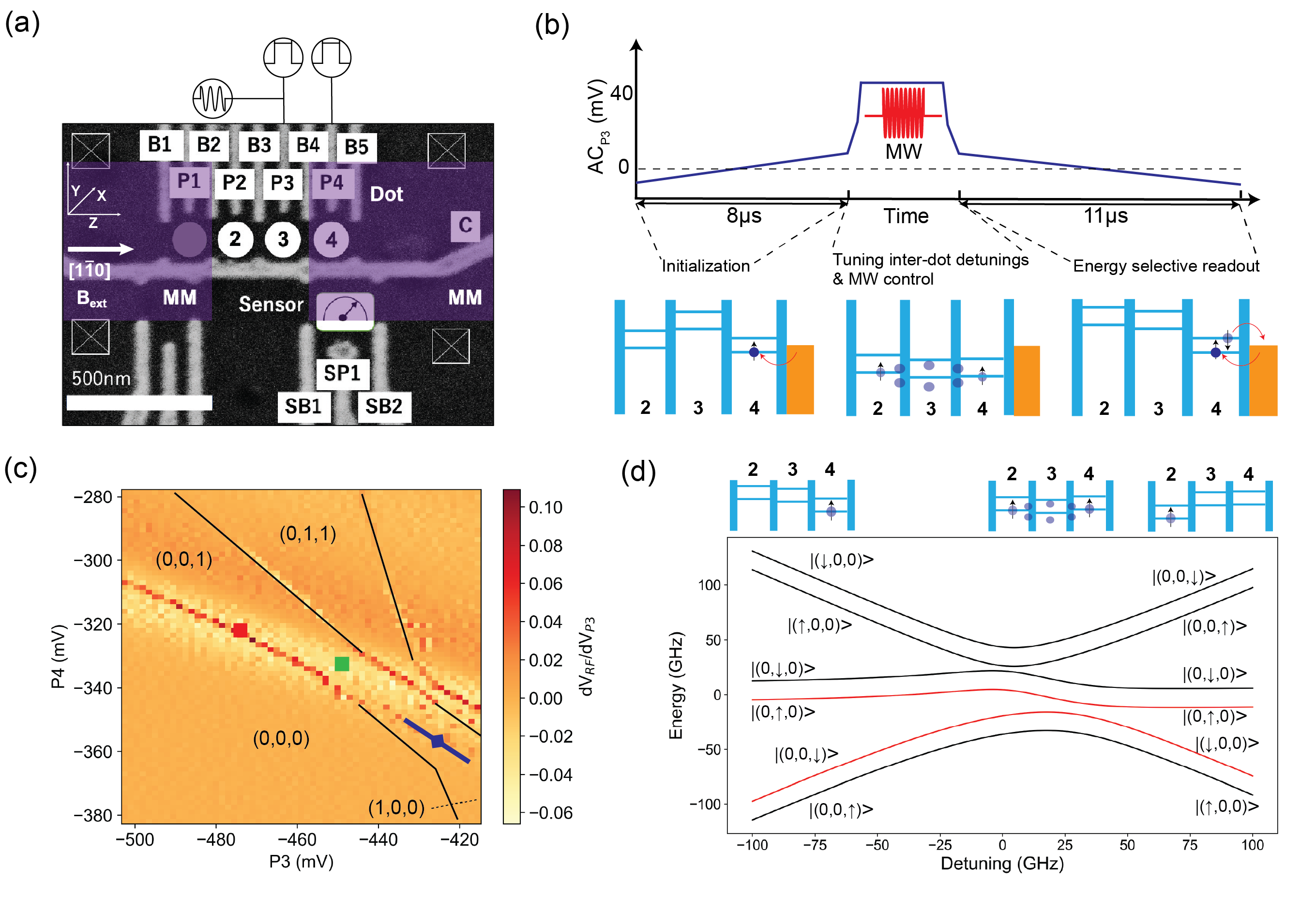}
\caption{\label{fig:wide}Characteristics of the single qubit control speed and coherence time:
(a)SEM picture of quadruple quantum dot device similar to the one used for this work. Blue squares indicate the fabricated micro-magnets. White squares indicate the ohmic contacts. 
(b) The pulse measurement scheme for single spin qubit control in TQD.
(c) Charge stability diagram of the triple dot used for this work. The red, green, and blue squares indicate the voltage value used for initialization\&readout, shuttling, and operation, respectively.
(d) Energy diagram of different charge and spin states in the TQD mainly used in the experiment as a function of global detuning ($\epsilon \equiv \tau_{z}^{(2-3)} + \tau_{z}^{(3-4)}$). The condition in which the tunneling couplings are sufficiently large relative to the Zeeman energy for stable spin manipulation. We calculated it with $2t_{23}$ = 16 GHz, $2t_{34}$ = 25 GHz, the Zeeman energy of 17 GHz,$b_{z}^{2-3}$ = 0.05 mT, $b_{z}^{3-4}$ = 0.001 mT, $b_{x}^{2-3}$ = 0.2 mT, $b_{x}^{3-4}$ = 0.01 mT, $\epsilon_{2-3}$ = 15 $\mu$eV, and $\epsilon_{3-4}$ = 15 $\mu$eV. We note that $\epsilon_{2-3} = 15 \mu$eV and $\epsilon_{3-4} = 15 \mu$eV at global detuning = 0.}
\end{figure*}

In this work, we investigate enhancing the DQD flopping-mode electron dipole spin resonance (EDSR) by optimizing TQD parameters. We explore if fine-tuning inter-dot orbital states in a TQD allows for the simultaneous optimization of Rabi frequency and coherence time. Next, we develop and demonstrate a feedforward neural network (FNN) approach that leverages past measurement data in the feedback cycle to achieve accurate qubit frequency estimation with fewer measurement points, improving upon conventional Bayesian methods for noise characterization in the flopping-mode regime. Finally, we combine this FNN-based feedback control system with pulse optimization technique to evaluate achievable gate fidelities in our GaAs device, where electron spins typically suffer from strong hyperfine interactions. 

\begin{figure*}
\centering 
\includegraphics[width=18cm]{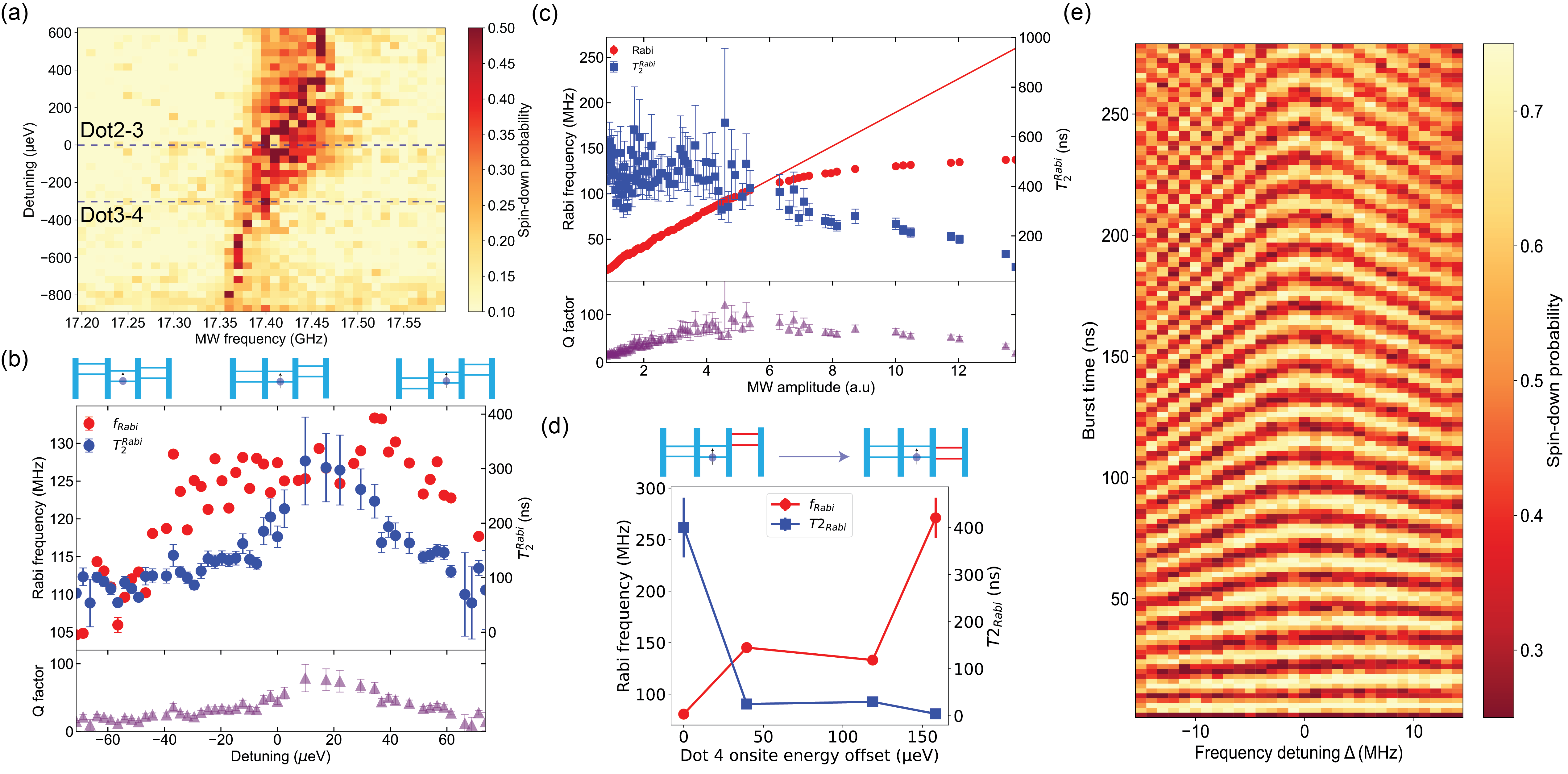}
\caption{\label{fig:wide} Dependence of the $f_{Rabi}$ and $T_{2}^\mathrm{Rabi}$ (a) The EDSR resonance spectroscopy as a function of inter-dot detuning between dot 2-3. The dashed lines indicate inter-dot transition between dot 2-3 and dot 3-4. (b)The top panel shows the dependence of $f_\mathrm{Rabi}$ and $T_{2}^\mathrm{Rabi}$ on inter-dot detuning of dot 2 and 3($\mu$eV). The bottom panel shows the corresponding Q factor. (c) The top panel shows the dependence of $f_\mathrm{Rabi}$ and $T_{2}^\mathrm{Rabi}$ on the MW amplitude. The bottom panel shows the corresponding Q factor. (d) The dependence of $f_\mathrm{Rabi}$ and $T_{2}^\mathrm{Rabi}$ on the onsite energy of dot3, that modulates the excited orbital energy level. (e) The Rabi oscillations as a function of MW frequency detuning at the spin sweet spot.}
\end{figure*}

We use a quadruple quantum dot (QQD) device fabricated on a GaAs/AlGaAs heterostructure (Figure 1(a)). An electron is confined in a multiple QD by the electrostatic potentials induced by voltages applied on Ti/Au gate electrodes. A pair of Co micro-magnets (MMs) is designed to create a large magnetic field gradient in $z$ direction ($>50$ mT) between adjacent dots for fast Rabi oscillations, while minimizing the magnetic field gradient along $z$-direction to maintain long coherence times. The MMs are fabricated on the surface after depositing the $\mathrm{Al_2O}_3$ insulator layer and magnetized by a magnetic field of $B_\mathrm{ext} = 2.5$ T applied in the $z$ direction. In this work, we focus on a single electron spin qubit confined in a triple QD (TQD) (dots 2, 3, and 4).

The measurement scheme is shown in Figure 1(b). When we control the single-spin state, the inter-dot detuning is pulsed to a certain value by arbitrary waveform generator (AWG), and then microwave (MW) pulse is applied to
perform EDSR. We initialize and measure the spin states using the ramped spin readout scheme\cite{ramp}, where we detect the spin state by reading the charge signal while gradually varying the voltage value so that only spin-down can tunnel out of the QD during readout. We chose the ramp rate such that the spin-up electron is adiabatically loaded into dot 4 during the initialization.
For implementing fast MW frequency feedback control, we develop an FPGA-based readout signal processing system. In this system, the demodulated readout signal is fed into the development board that equips the FPGA, which performs threshold processing, and the resulting single-shot outcomes (0 or 1) are stored in registers. These single-shot outcomes stored in the registers are then processed by the FNN, and the NCO (Numerically Controlled Oscillator) frequency is updated according to the estimated frequency detuning.

Figure 1(c) shows the stability diagram of the TQD measured.
We initialize and readout at position I (the red square) of dot 4, and then, move closer to the target detuning point adiabatically (the green square), then control inter-dot detuning of 2-3 along the O line segment (the blue square and line), and perform EDSR. Then the electron is shuttled to dot 4 again and readout at position R. Since the tunnel coupling between dots 2$t_{23}$ ($\sim$16 GHz) is adjusted so that $z$-component of the magnetic field gradient $b_{z}$ is as small as possible, the range of sweet spot becomes larger than the amplitude of the charge noise and and $b_{x}$ becomes relatively large. By satisfying these conditions,a sweet spot can be created and the Rabi frequency and coherence time are maximized at the same point.
In addition, dot 4 is very strongly coupled to dot 3($2t_{34}$ $\sim$ 25 GHz). This additional dot allows to tune precisely the energy differences and the gradient of the inter-dot excited orbital levels. 

\section{\label{sec:Results}Results}

\subsection{The characteristics of the single spin qubit}
First, we explain the QD conditions to achieve high-fidelity qubit control in our GaAs device by maximizing Rabi frequency and coherence time. For this purpose, we use states of which wavefunction spread over the dot 2, 3, and 4 to extend the controllability of flopping-mode EDSR\cite{flopping,flopping2}.

In the presence of a micro-magnet and an external magnetic field, the Hamiltonian for a single electron in a triple quantum dot (TQD) can be written as,
\begin{equation}
H = H_{c} + H_{s}.
\end{equation}
The charge and spin terms are described by,
\begin{equation}
H_{c} = \sum_{\langle i,j \rangle} \left( \frac{\epsilon_{ij}}{2} \tau_{z}^{(ij)} \right) + \sum_{\langle i,j \rangle} t_{ij} \tau_{x}^{(ij)},
\end{equation}
\begin{equation}
H_{s} = \frac{1}{2} g\mu_{B} \left[ B_Z \sigma_{z} + \sum_{\langle i,j \rangle} \left( b_x^{(ij)} \sigma_{x} + b_z^{(ij)} \sigma_{z} \right) \tau_{z}^{(ij)} \right],
\end{equation}
where $\epsilon_{ij}$ is the detuning between dot $i$ and $j$, $t_{ij}$ is the inter-dot tunnel coupling between dots $i$ and $j$, $\tau_{z}^{(i)}$ and $\tau_{x}^{(ij)}$ are Pauli operators in position space for dot $i$ and between dots $i$ and $j$, respectively. $\sigma_{i}$ are the Pauli operators for the spin degree of freedom, $B_{Z}$ is the external magnetic field in $z$ direction, and $b_{x}^{(ij)}$ and $b_{z}^{(ij)}$ are the differences in the magnetic field between dot $i$ and $j$ in the $x$ and $z$ directions caused by the micro-magnet.

Thus, the dynamics of a single electron spin in a TQD can be described by the multi-level system shown in the energy diagram of Figure 1(d), where the solid lines show the TQD eigenenergies with spin-up and spin-down components. The energy diagram is ploted as a function of global detuning. Here, the global detuning is defined as $\epsilon \equiv \tau_{z}^{(2-3)} + \tau_{z}^{(3-4)}$ in accordance with the notation in Eq. (2). The inter-dot tunnel couplings $t_{ij}$ lead to anti-crossings near specific detuning points. The transverse magnetic field gradients $b_x^{(ij)}$ from the micro-magnet hybridize the spin and orbital states near these anti-crossings, while the longitudinal gradients $b_z^{(ij)}$ induce different energy splittings between states in the largely-detuned regions in the TQD. 

Spin-charge hybridization is maximized when the excited inter-dot orbital states are closest to the ground state, resulting in the fastest Rabi frequency. In Fig. 1(c), the two excited orbital levels (indicated by red lines) are the primary targets for fine-tuning this hybridization. In particular, the transverse magnetic field component, $b_{x}^{(ij)}$, induces a pronounced dip in the resonance frequency near the anti-crossing owing to spin-charge hybridization\cite{Studenikin2019_spincharge}. When the effect of $b_{x}^{(ij)}$ exceeds that of the longitudinal component $b_{z}^{(ij)}$, the center of this dip coincides with a sweet spot\cite{flopping,flopping2}. However, as $b_{z}^{(ij)}$ increases, the sweet spot shifts away from the zero-detuning point and may eventually disappear, leading to a monotonic variation of the resonance frequency with detuning. Moreover, the sweet spot must be sufficiently broad compared to the standard deviation of the detuning noise (i.e. $2t_{23} \gg \sigma_{\text{detuning}}$). To maximize qubit performance, it is therefore essential to optimize the Hamiltonian parameters so that both the Rabi frequency and the coherence time are maximized at the same detuning point in the TQD system.

Next, we show the results of the performance of the flopping-mode EDSR in the TQD.
Figure 2(a) shows the EDSR spectroscopy as a function of inter-dot detuning between dot 2-3. The two dashed lines indicate the inter-dot transitions between dot 2-3 and dot 3-4. After a monotonic transition of the resonance frequency from dot 4 to dot 3, from dot 3 to dot 2, the resonance width between dots 2-3 increases rapidily as it approaches zero detuning. Simultaneously, the shift in resonance frequency remains comparatively small in this region as a function of inter-dot detuning. (see supplemental information). This combination, where the resonance width (which depends on $b_x^{23}$) grows significantly while the resonance frequency changes minimally, creates an operating point where the system is less sensitive to charge fluctuations (sweet spot).

The top panel of Fig. 2(b) shows the Rabi frequency $f_\mathrm{Rabi}$ (red), and Rabi decay time $T_{2}^\mathrm{Rabi}$(blue) as a function of the inter-dot detuning of the center DQD (dot2-3) at 10 dBm MW power. Both are maximized near zero detuning point between dots 2 and 3. The small offset of the maximum $T_{2}^\mathrm{Rabi}$ from zero detuning reflects $b_z^{(ij)}$ between dots 2 and 3\cite{flopping}. As a result, the Q factor (purple ) takes maximum near the zero detuning point as shown in the bottom panel of Fig. 2(b). Overall trend of the Rabi frequency can be understood in analogues to flopping-mode EDSR\cite{flopping2}. While in the flopping-mode EDSR in DQD systems, the Rabi frequency can be expressed as $f_\mathrm{Rabi}\propto2\Omega g \mu_B b_x /\left|4 \Omega^2-E_z^2\right|$, where $\Omega$ = $\sqrt{\varepsilon^2+4 t_{ij}^2}$ and $E_z = g\mu_{B}B_Z$, the addition of dot 4 with strong tunnel coupling ($t_{34} >> t_{23}$) should modify the standard flopping-mode EDSR behavior. The strong coupling to dot 4 imposes another large aniti-crossing to the energy diagram of the center DQD as shown in Fig. 1(c), allowing us to tune both the energy levels and effective inter-dot spin-orbit coupling between dots 2 and 3 through the detuning between dots 3 and 4.
This additonal tunability via dot 4  would also enable to tune the dependence of the resonance frequency of the flopping-mode qubit in the center DQD  on the detuning between dots 2 and 3 smaller. This makes the system more robust against quasi-static noise\cite{flopping2}, leading to improved coherence times.

In addition, we optimize the MW power to maximize the Q factor, which is defined by $T_{2}^\mathrm{Rabi}$/2$f_\mathrm{Rabi}$, at the optimal detuning point ($\epsilon$ = 15$\mu$eV). Fig. 2(c) shows the MW power dependence of $f_\mathrm{Rabi}$ (red), $T_2^\mathrm{Rabi}$ (blue), and Q factor (purple). $f_\mathrm{Rabi}$ linearly increases to 100 MHz and saturates to around 140 MHz. No significant decrease in $T_{2}^\mathrm{Rabi}$ occurs in the linearly increasing regions, whereras it decreases gradually in the saturated regions, possibly due to heating of the device. 

To check the controllability of energy levels by tuning the energy of dot 4, we demonstrate an additional degree of control over the qubit performance through the onsite energy of the third dot (dot 4). Maximizing the Rabi frequency primarily requires matching the excited state splitting ($E_{\text{excite}} - E_{\text{ground}}$) to the Zeeman energy ($E_z$), this can be achieved through its energy level tuning of dot 4. Fig. 2(d) shows how adjusting the onsite energies of dot 4 at zero detuning (with MW power of 10 dBm) affects both $f_\mathrm{Rabi}$ and $T_{2}^\mathrm{Rabi}$. By bringing the excited orbital states closer in energy through this third-dot tuning, $f_\mathrm{Rabi}$ reached to 271 ± 20 MHz - comparable to those reported in Ge-based nanowire systems\cite{Gefast1,Gefast2}. However, we observe a trade-off: while closer orbital spacing enables faster Rabi oscillations, it also increases sensitivity to charge noise, leading to decreased $T_{2}^\mathrm{Rabi}$. When the flopping-mode qubit in the center DQD is concerned, this demonstrate that dot 4 serves an additional control parameter in the speed-coherence trade-off.

Fig. 2(e) shows the Rabi oscillations as a function of the MW frequency using MW power of 4 dBm (MW amplitude of 5). The combination of fast Rabi oscillations and operation at the spin sweet spot results in a well-defined chevron pattern, demonstrating the system's resilience against nuclear spin and charge noise. The asymmetry in the chevron pattern between positive and negative MW frequency detuning from the resonance suggests an inherent asymmetry in the frequency response of the triple-dot system. We speculate that this asymmetry originates from different frequency characteristics between the primary excited inter-dot orbital states mediating the spin rotation and the second excited inter-dot orbital state within the triple-dot charge eigenstate manifold. Although the frequency dependence of the MW amplitude could contribute to this behavior, the observed asymmetry with respect to detuning suggests that the influence of second excited orbital states is a plausible explanation.

Compared with previous experimental demonstrations on the flopping-mode qubit\cite{flopping,flopping4}, we specifically design the magnetic field gradient of MM for flopping-mode operation, optimize the tunnel couplings and the inter-dot orbital excited states using a triple dot configuration to maximize the Q factor, and demonstrate the simultaneous enhancement of $f_\mathrm{Rabi}$ and $T_{2}^\mathrm{Rabi}$ using the flopping-mode qubit for the first time. 

\subsection{Extending the coherence time using measurement based feedback control}
To characterize and mitigate the influence of low-frequency noise on the frequency of the flopping-mode qubit at the sweet spot(see Fig. 2(e)), we use measurement-based feedback operation of the qubit control microwave\cite{feedback,feedback2}. This approach offers advantages over decoupling pulses\cite{echo,CPMG,CPMG2} as it requires only additional measurements without modifications to the target-qubit operation. While Bayesian estimation is commonly used for resonance frequency estimation\cite{adap,feedback2}, we implement a feedforward neural network (FNN) approach for improved stability and accuracy with fewer measurements. By leveraging information from past Ramsey measurements and incorporating temporal noise correlations\cite{predict,predict2}, our approach enables efficient estimation even with sparse sampling that omits shorter free evolution times ($t_R$)\cite{adap}, significantly reducing the total measurement overhead (see Supplemental information).

\begin{figure*}
\centering
\includegraphics[scale=0.36]{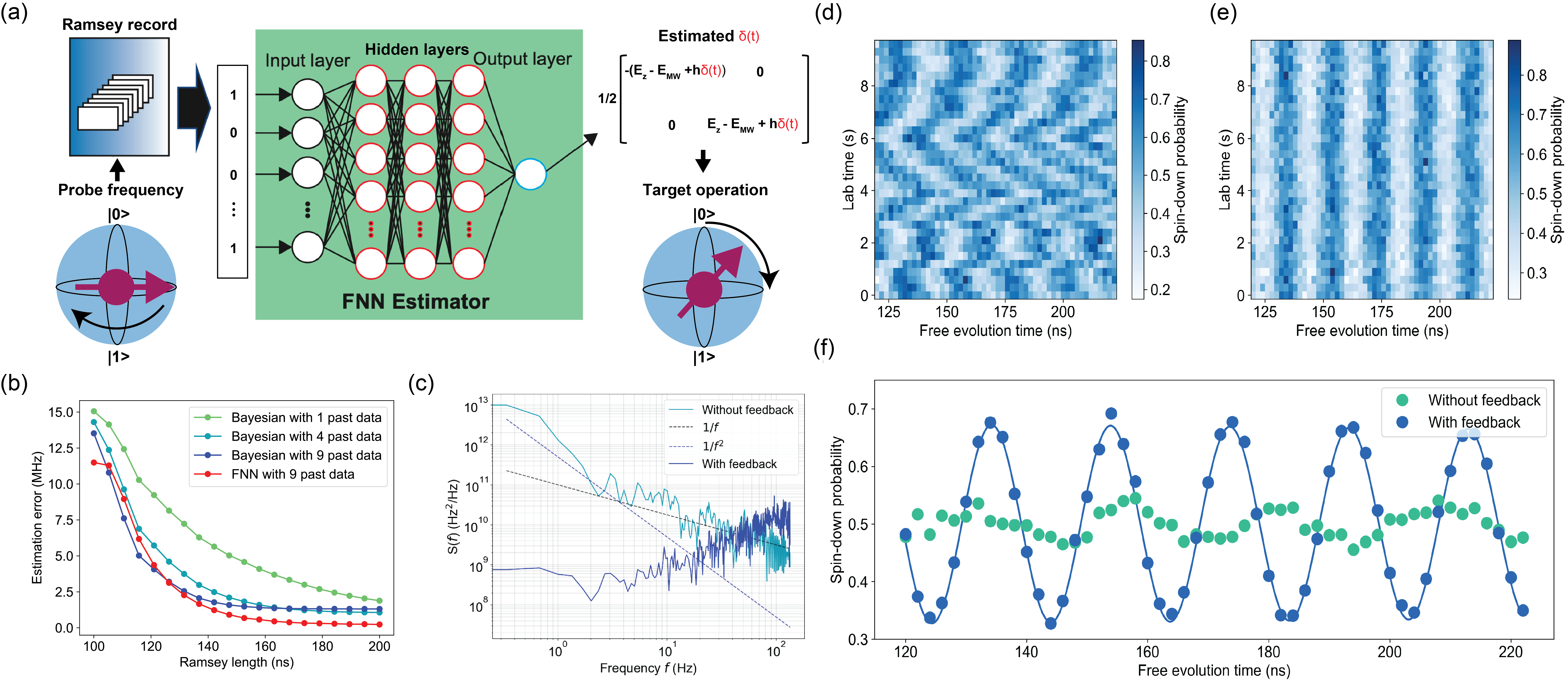}
\caption{\label{fig:wide} Performance of the measurement based feedback control of spin qubits (a) The scheme of the FNN based feedback control of single-spin operation. (b) Simulated performance of the qubit frequency estimation by the Bayesian and the FNN method as a function of the maximum wait time of Ramsey measurement (Ramsey length). (c) Noise power spectrum density at the qubit frequency with (Blue) and without (Cyan) feedback control of MW at sweet spot. Black and blue dashed lines indicate 1/$f$ and 1/$f^{2}$ trend. (d) Measured spin-down probability as a function of free evolution time $\tau$ between two $\pi$/2 pulses with equal phase. The oscillation frequency fluctuates in time. (e) The same measurement as in (d) but obtained with the feedback control. We update the MW frequency according to the estimated $\delta(t)$. (f) The averaged Ramsey oscillations over (a)(cyan) and (b) (blue).}
\end{figure*}

As shown in Fig. 3(a), our FNN-based feedback system processes single-shot Ramsey measurement outcomes ($m_1,m_2...m_k$) from past measurements using two hidden layers (300 neurons each) with sigmoid activation functions. This architecture allows probabilistic information to propagate similarly to Bayesian posterior distributions, with the output layer estimating resonance frequency shift $\delta(t)$.

We trained our model on simulated data incorporating experimental noise characteristics. After extracting Rabi frequency and visibility from initial measurements, we generated training data from simulated Ramsey oscillations and used the noise power spectrum density from previous measurements\cite{feedback} to produce realistic frequency drift patterns. The FNN parameters were optimized by analyzing estimation error across various configurations (see Supplemental information).

Fig. 3(b) compares the estimation errors between FNN and Bayesian methods as a function of Ramsey length, that is, measurement number in the simulations. We evaluated mean absolute error across 100,000 simulated datasets (using $\alpha=0.32$ and $\beta=0.23$ in our Ramsey oscillation model; see Supplemental information). Measurements start at 100 ns with 1 ns increments, each providing a single binary outcome. The error rate of the Bayesian estimation with 10 past datasets decreases rapidly with fewer measurement points, and then saturates as more data points are added. This saturation indicates a loss of temporal correlation information, as the estimation becomes effectively equivalent to that obtained by averaging over 0 past datasets. In contrast, the FNN maintains estimation accuracy by optimally weighting past data and preserving information from early fluctuations.

Next, we analyze the noise power spectrum density (PSD) at the qubit frequency using repeated Ramsey oscillations and FNN estimation\cite{feedback} (Fig. 3(c)). Our protocol uses 62 single-shot measurements with varied intervals to estimate frequency detuning, then updates MW frequency based on FNN output. The PSDs with and without feedback control (blue and cyan, respectively) reveal different noise characteristics. Above 2 Hz, we observe a 1/$f$ dependence, indicating that flopping-mode qubit coherence in our GaAs QDs is primarily limited by charge noise via MM field gradients rather than nuclear spin noise\cite{feedback}. Below 2 Hz, the spectrum transitions toward a 1/$f^2$ trend, suggesting increased nuclear spin noise contribution at lower frequencies\cite{noise}. Feedback operation significantly suppresses low-frequency components. The slight increase in high-frequency components can be attributed to the estimation errors and measurement delays.

Finally, we demonstrate the feedback control of the MW frequency using the trained FNN. We collect 62 single-shot data points from Ramsey oscillations with intervals $t_R=100,102, \ldots 222 \mathrm{~ns}$ for each estimation. Each measurement cycle ($T_\mathrm{shot}=22.0$ $ \mu \mathrm{s}$) includes initialization (8$\mu s$), spin manipulation (1$\mu s$), readout (11 $\mu s$), and AWG pulse compensation (2 $\mu s$), resulting in the estimation time, $T_{est}=62 t_R \approx 1.36 \mathrm{~ms}$ per estimation. Additionally, the NCO frequency update time of 2 ms is added to this estimation time. We continuously update $f_{\text {qubit }}^{\text {est }}$ based on the FNN-estimated $\delta(t)$ and adjust the microwave frequency to $f_{\mathrm{MW}}=f_{\text {qubit}}^{\text {est}}+\delta(t)$. Comparing Ramsey oscillations with and without feedback control (Fig. 3(d) and (e)), we observe significant suppression of period fluctuations and improved coherence.

Fig. 3(f) shows averaged Ramsey oscillations from both conditions. Without feedback, $T_{2}^{*}$ is only 52.42 ± 2.13 ns from direct fitting (see supplemental information). Due to sparse sampling in our protocol, we analyze frequency fluctuations in the Ramsey oscillations with feedback rather than direct fitting. Using the relation $T_{2}^{*} = \sqrt{2}/(2\pi\sigma_f)$\cite{feedback} where $\sigma_f$ is the standard deviation of frequency fluctuations, we obtain $T_{2}^{*}$ = 580 ± 10 ns. Notably, our flopping-mode architecture maintains the extended coherence times while operating at Rabi frequencies exceeding 100 MHz, demonstrating robust resilience to low-frequency noise at fast operation speeds. Further improvements in readout parameters could reduce estimation periods and suppress remaining high-frequency noise contributions.

\subsection{The evaluation of the qubit performance}
To validate a high fidelity single qubit operation, we characterize the overall single qubit control fidelity using randomized benchmarking (RB). We use a Clifford gate set containing $\pi / 2$ rotations around the Bloch sphere (see Supplemental information). Fig. 4(a) shows the RB measurement protocol with the feedback cycle. After measuring single-shot outcomes with 62 different free evolution times (100 ns, 102 ns...222 ns), we perform the RB at a certain number ($N$) of Clifford gates with an updated MW frequency. 

In addition, we employ piecewise constant (PWC) optimization of the pulse $\pi /2$ to further improve control fidelity\cite{PWC,PWC2}. Although this method is used primarily to find pulses that suppress unwanted resonances, in our case, by the second inter-dot excited state, especially for short-duration pulses\cite{PWC,PWC2}, it may also be beneficial in mitigating heating effects as observed in \cite{opt}. We employ the covariance matrix adaptation-evolution strategy (CMA-ES) algorithm to optimize the large number of parameters of the $\pi$/2 pulse\cite{CMA}. We optimized the in-phase (I) and quadrature (Q) amplitude components at 4.05 ns with a population of 20 different pulse configurations. The optimization was performed over 40 iterations, and we used the best-performing configuration among all iterations. The cost function was defined as the sequence fidelity at sequence length 5 in randomized benchmarking (RB), as described in Ref.~\cite{PWC}.

Fig. 4(b) and (c) show the I and Q amplitude components of the pulse envelope before (black) and after PWC optimization (red, blue), respectively. The duration of the $\pi /2$ pulse $\tau_g$ is 4.05 ns.
Fig. 4(d) shows the RB results with and without feedback operation in the sweet spot region. As the number of applied Clifford gates $m$ increases, the fidelity of the standard sequence decays to $F(m)$, with the depolarizing parameter $p_\mathrm{c}$. Then, the average single gate fidelity $F_{\text {average}}$ is calculated using the following equation\cite{takeda},
$F_{\text {average }}=\frac{1+p_{\mathrm{c}}^{\text {single}}}{2} \sim 1-\frac{1-F_{\mathrm{c}}}{N_{p}}$,
where $p_{\text {single}}=\left(p_{\mathrm{c}}\right)^{\frac{1}{N_{p}}}$ represents the decrease of the sequence fidelity per single primitive gate and $N_{p}$ is the average number of primitive gates per one decomposed Clifford gate. In our case, $N_{p}$ = 3.217.
Since our gate set consists only of $\pi /2$ gates, the average single-qubit control fidelity corresponds to the $\pi /2$ gate fidelity $F_{\pi /2}$. We observe an increase in $F_{\pi /2}$ from 99.23$\pm$0.12$\%$ to 99.56$\pm$0.03$\%$ for feedback control and to 99.72$\pm$0.18$\%$ for optimized PWC pulse. 

The improvement in the fidelity through the feedback control directly reflects the suppression of frequency fluctuations observed in the PSD measurements. For a $\pi/2$ rotation around the $x$-axis with $\tau_g = 4.05$ ns, the measured RMS frequency fluctuation of $\sigma_f \approx 3.87$ MHz without feedback leads to a gate error of $\epsilon_\text{err} = \frac{1}{6}(2\pi\sigma_f \tau_g)^2 \approx 0.16\%$. Our FNN-based feedback reduces $\sigma$ to 388 kHz, as measured from the frequency noise suppression shown in Fig. 3(f), decreasing the error to 0.0016\%. This predicted improvement agrees well with the observed fidelity increase from 99.23\% to 99.56\%. The further enhancement to 99.72\% through PWC pulse optimization stems from both the correction of the frequency spectrum\cite{Rimbach-Russ_2023} to suppress transitions to the second inter-dot excited state and the improved robustness against heating effects\cite{opt}. The remaining infidelity is likely limited by high-frequency charge noise that cannot be compensated by our feedback scheme due to the 3.36 ms estimation delay.

\begin{figure*}
\centering
\includegraphics[scale=0.5]{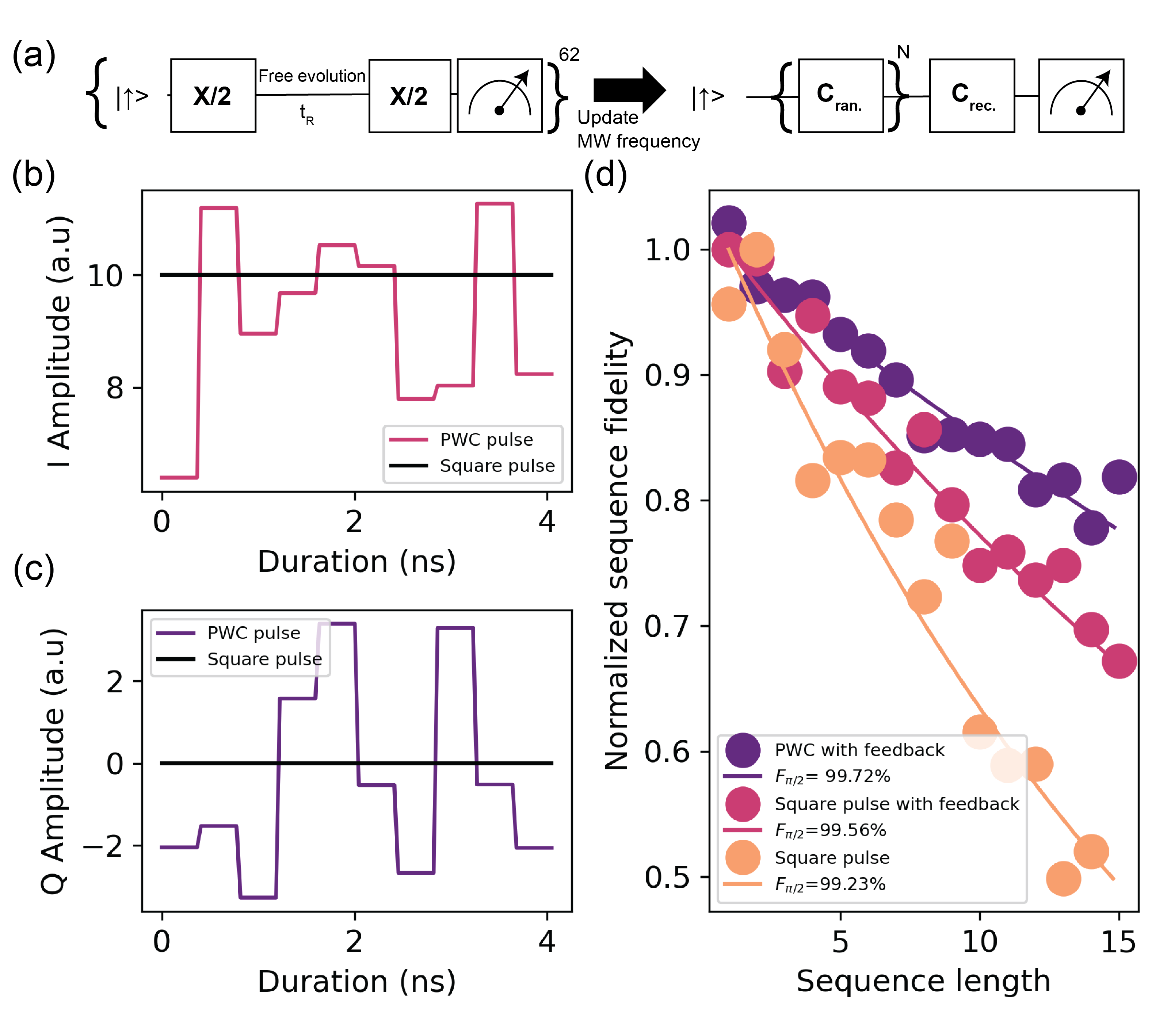}
\caption{\label{fig:wide} Single qubit fidelity evaluation using randomized benchmarking (a) Measurement protocol of RB with the feedback cycle. After measuring single-shot outcomes with 62 different free evolution time (100 ns, 122 ns...222 ns), we perform RB at a certain number (N) of Clifford gates with the updated MW frequency. (b),(c) show in-phase and quadrature amplitude component of the pulse envelope before (black) and after the piecewise-constant optimization (red, blue), respectively. (d) Standard RB at the sweet spot region. Each data point is an average over 30 randomly chosen sequences of the respective length. We normalize the state probability to remove the readout error.}
\end{figure*}
\section{Conclusion}

In this work, we extended the flopping-mode developed in DQD to a TQD and showed that the flopping-mode with TQD configuration enables precise control over orbital energy levels, allowing us to achieve Rabi frequencies exceeding 100 MHz while maintaining coherence through proper tuning of the inter-dot orbital states. This demonstrates that device-specific parameters can be effectively utilized for high-fidelity control, rather than relying solely on material properties\cite{Gefast1,Gefast2}.

Moreover, we implemented a machine learning-based feedback control system that innovates on conventional Bayesian methods by incorporating temporal correlations through multiple past Ramsey measurements. We found that our FNN-based approach has more accurate frequency estimates with fewer single-shot measurements, while providing detailed insights into the device's noise mechanisms, revealing a transition from charge noise to nuclear spin noise dominance at lower frequencies\cite{noise}. This combination of efficient noise characterization and mitigation represents an important protocol for optimizing quantum dot devices.

Finally, through randomized benchmarking, we demonstrated that a $\pi$/2 gate fidelity of 99.72\% with a gate time of 4.05 ns was achieved by combining feedback control with piecewise constant pulse optimization\cite{CMA}. This result validates our comprehensive approach to qubit control optimization through device engineering, parameter tuning, and active noise suppression.

Our work provides a generalizable framework for achieving high-fidelity spin control in semiconductor quantum dot arrays by utilizing device-specific parameters rather than material dependent properties or external field gradients. The combination of engineered inter-dot orbitals and machine learning-based noise characterization and control techniques demonstrated here can be readily extended to other multiple quantum dot platforms, including Si and Ge systems\cite{six,Ge2,John2024TwoDimensional}, providing a pathway toward scalable quantum computation with spin qubits.

\section{Methods}
\subsection{Experimental setup}
The device is fabricated on a GaAs/AlGaAs heterostructure, where a two-dimensional electron gas (2DEG) is formed at a depth of 100~nm below the surface. Fine gate electrodes consisting of Ti (10~nm) and Au (25~nm) are defined by electron-beam lithography and deposited by evaporation. After gate deposition, a 50~nm-thick Al$_2$O$_3$ insulating layer is grown by atomic layer deposition (ALD). Subsequently, a 250~nm-thick micro-magnet is deposited on top of the insulating layer.
The device is cooled in a dry dilution refrigerator (Bluefors) to a base electron temperature of around 170 mK. The DC voltage applied to the gate electrodes is generated by QDAC. Control pulses for manipulating the charge states are generated by an arbitrary waveform generator (Tektronix 5014B). Microwave signals are generated by microwave signal generators (Keysight N5173). The microwave signal is modulated by the IQ mixer (500-900 MHz) to prevent unintentional spin rotations due to microwave carrier leakage. Radiofrequency reflectometry is used for fast measurement of the charge sensor conductance. The right reservoir of the sensor dot in Fig. 1(a) is connected to a tank circuit with an inductance of 1.2 $\mu$H and a resonance frequency of 148 MHz. The reflected signal is amplified, demodulated and digitized using a digitizer (Spectrum M2p). Feedback control of MW frequency is implemented by the Zynq UltraScale + RFSoC ZCU111 Evaluation Kit (AMD). We investigated multiple platforms for implementing the Hamiltonian parameter estimation, including both GPU and FPGA implementations. The GPU implementation achieves processing times of approximately 100 $\mu$s, while the FPGA implementation demonstrates faster processing at around 3 $\mu$s. Despite the FPGA's superior processing speed, we ultimately chose to primarily utilize the GPU for estimation. This decision is driven by three primary considerations: the FPGA-implementable model size is restricted to networks with only two hidden layers of up to 100 neurons each; the FPGA implementation would require regenerating the FPGA image whenever the model needed to be updated; and the overall feedback timing is fundamentally limited by NCO frequency updates of approximately 2 ms. 

\section*{Acknowledgements}
This work was supported by JSPS KAKENHI (Grant No. 17H06120, 23H05458, 23H05455), JST Moonshot R\&D (Grant No. JPMJMS2066, JPMJMS226B), NRC Challenge Program (QSP-013), and the Dynamic Alliance for Open Innovation Bridging Human, Environment and Materials. A.L. and A.D.W. appreciate the support of DFG-TRR160 and BMBF Q.Link.X (Grant No. 16KIS0867). T.M. appreciate the support of JST (Grant No. JPMJPF2014) and MEXT Q-LEAP (Grant No. JPMXS0120319794).

\section*{Author contributions statement}
YM conceived these experiments. YM and XL conducted these experiments. KK, TM, and YM develop the FPGA image for feedback control. YM analyzed the results. AL and ADW conceived the heterostructure and grew the layers by molecular beam epitaxy. YM fabricated the device.  YM, TF, and AO interpret and discuss the measurement results. AO motivated, led the base, surveyed and organized these investigations. All authors reviewed the manuscript. 
\section*{Data availability}

\section*{Competing interests}
The authors declare no competing interests.

\section*{\label{sec:Supp}Supplemental information}

\subsection{Micro-magnet simulation}

We design the MMs to get a large difference of magnetic field components in the $x$ direction between the left and right dots $b_x$ for fast Rabi frequency and a small difference of magnetic field components in $z$ direction between left and right dots $b_z$ for a long coherence time in the center DQD. The simulated longitudinal and transverse magnetic fields are shown in Fig. S1(a) and S1(b), respectively.
Assuming the distance between two dots of the center DQD as 100$\sim$200nm, $b_x$ for the inter-dot tunneling in the center DQD is higher than 50 mT in this MM design. $b_z$ is zero if the positions of each QD in the center DQD are symmetrical to the magnetic field distribution of the MMs, but as it becomes asymmetrical, a difference of up to 100 mT is created. This is actually a potential problem for qubit coherence, and indeed we sometimes observe the large $b_z$ and short coherence time in experiment. We solve this problem by probing and adjusting $b_z$ using EDSR resonance frequency spectroscopy (Fig. S2).

\setcounter{figure}{0}
\renewcommand{\figurename}{FIG.}
\renewcommand{\thefigure}{S\arabic{figure}}
\begin{figure*}[ht]
\centering
\includegraphics[width=1\textwidth]{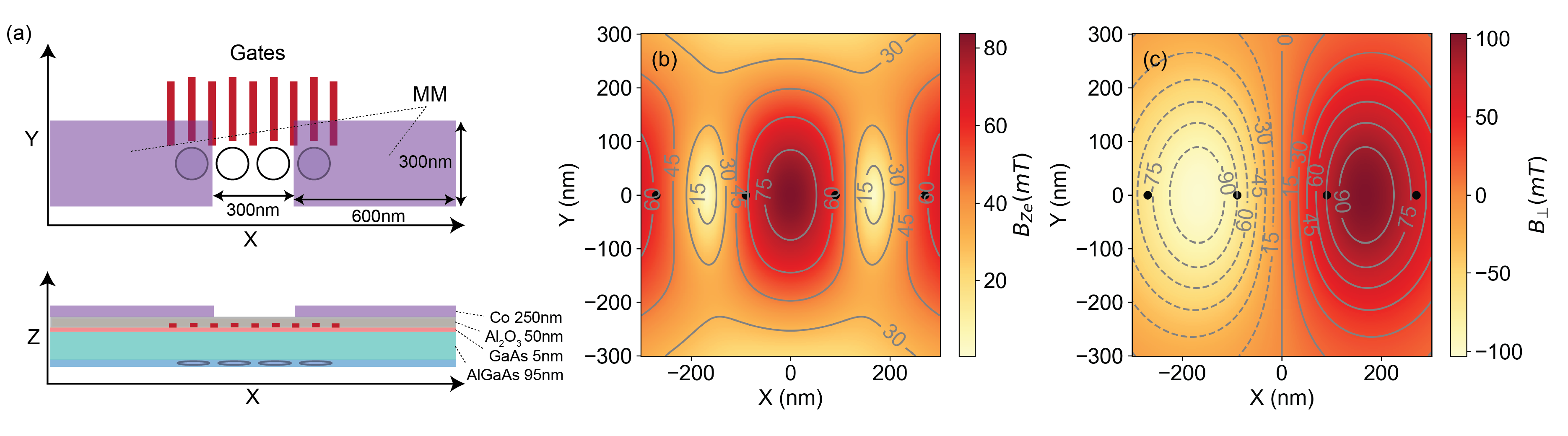}
\caption{\label{supp:fig:MM_sim} \textbf{Simulated magnetic field induced by micro-magnets.}
a) Schematic diagram of the micro-magnet design. White circles indicate the intended positions of quantum dots, red rectangles represent gate electrodes, and purple rectangles represent micro-magnets. The top figure shows dimensions in the X-Y plane, and the bottom figure represents the cross-sectional view in the X-Z plane.
b) The simulated transverse magnetic field. The dot positions are indicated by black dots. c) The simulated longitudinal magnetic field. We assume the magnetization of 1.8T for the simulation.
}
\end{figure*}

\begin{figure*}[ht]
\centering
\includegraphics[width=0.6\textwidth]{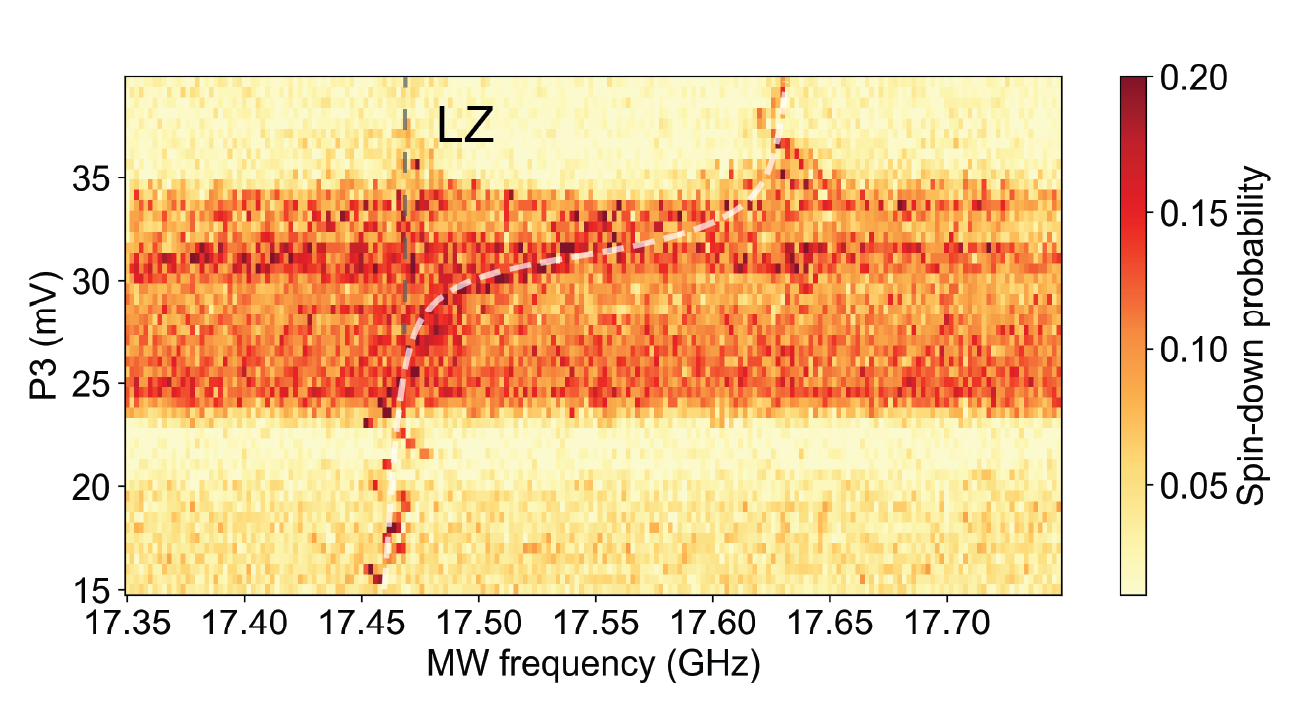}
\caption{\label{supp:fig:MM_sim} \textbf{Inter-dot condition with smaller tunnel coupling and very large $b_z$ } EDSR resonance scan as a function of the voltage of P3 (mV) and MW frequency (GHz). The white dashed line shows the shift of the resonance frequency. The black dashed line shows the remaining resonance frequency in dot 4 that is induced by Laudau-Zener excitation of charge. The high spin probability offset around the center voltage can be attributed to either thermal spin randomization or charge state leakage induced by microwave (MW) excitation.
}
\end{figure*}

\subsection{Comparison between Bayesian and FNN estimation in experiment}
In this section, we show the additional characterization results of the estimation algorithm we use in this work.

First, we explain the common method for estimating the qubit frequency of Ramsey oscillations based on a Bayesian approach\cite{adap,feedback2}, where the theoretical model is used to calculate the maximum likelihood of Hamiltonian parameters. 
In the Bayesian approach for real-time estimation, the longer the evolution time, the more sensitivity is gained by leveraging the accuracy gains from previous measurements, allowing estimations that exceed the standard limits for repeated measurements with a single evolution time. Let us denote the outcome of the $k$th measurement as $m_k$, which can be either $|\uparrow\rangle$ or $|\downarrow\rangle$. We define $P(m_k|\Delta\omega)$ as the conditional probability of obtaining measurement outcome $m_k$ given a frequency detuning $\Delta\omega$. This probability can be expressed as:
\begin{equation}
P(m_k|\Delta\omega) = \alpha(p_{\downarrow}) + \beta
\end{equation}

where $\alpha$ and $\beta$ are parameters determined by measurement error. In this formulation, we incorporate the experimentally determined Rabi frequency $\omega_\mathrm{Rabi}$ and $\pi/2$ pulse duration $t_{\pi/2}$.
Under the assumption of measurement independence, where previous measurement outcomes do not influence subsequent ones, we can express the conditional probability for $\Delta\omega$ given $N$ consecutive measurements as:
\begin{equation}
P(\Delta\omega|m_N, \ldots m_1) = \prod_{k=1}^N P(\Delta\omega|m_k),
\end{equation}
Applying Bayes' theorem, which states that $P(\Delta\omega|m_k) = P(m_k|\Delta\omega)P(\Delta\omega)/P(m_k)$, we can reformulate the above expression as:
\begin{equation}
P(\Delta\omega \mid m_N, m_{N-1}, \ldots, m_1)
=\mathcal{N}\,\left[\prod_{j=1}^{t} P_j(\Delta\omega)\right]
\left[\prod_{k=1}^{N} p_{\downarrow}(\Delta\omega)\right],
\end{equation}
where $\mathcal{N}$ represents a normalization constant, $P_j(\Delta\omega)$ denotes the prior distribution from $j$ steps ago (with $j=1$ corresponding to the most recent prior), and $t$ is the total number of prior distributions incorporated into the analysis. The estimation procedure involves measuring spin states for each measurement $m$ and subsequently identifying the value of $\Delta\omega$ that maximizes the posterior distribution $P(\Delta\omega|m_N, m_{N-1}, \ldots m_1)$.

\begin{figure*}[ht]
\centering
\includegraphics[scale = 0.55]{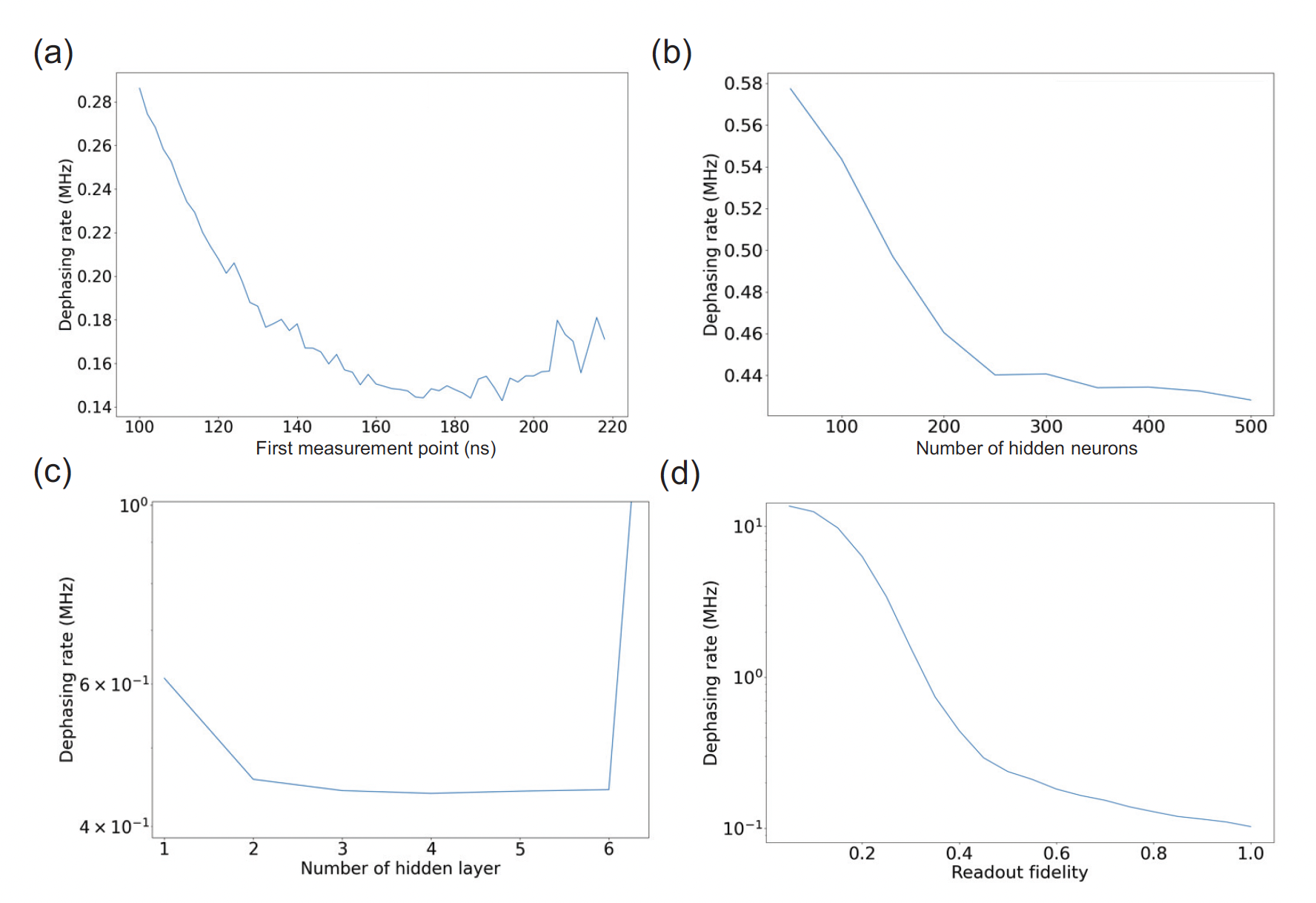}
\caption{Evaluation of the expected dephasing rate from the FNN based estimation as
a function of (a) the free evolution time of the Ramsey measurement for the initial measurement point, (b) the number of
neurons for each hidden layer, (c) the number of hidden layers in FNN, and (d) the readout
fidelity. The mean absolute error between the estimated and actual resonance frequencies
for the 100,000 datasets generated by simulated Ramsey oscillations is evaluated as the
estimation error.}
\end{figure*}

Fig. S4 shows the comparison of qubit frequency deviation estimates using: (a) Bayesian estimation with 1 past measurement, (b) 2 past measurements, (c) 9 past measurements, and (d) FNN with 9 past measurements.
As we expected from the simulation results, Bayesian estimation shows that while using past Ramsey data as prior distribution reduces estimation error, it gradually filters out high-frequency noise, leading to smoother results. On the other hand, when using FNN, even with 9 past data points as input, it retains sensitivity to high-frequency fluctuations without such filtering effects. 
\begin{figure*}[ht]
\centering
\includegraphics[scale = 0.6]{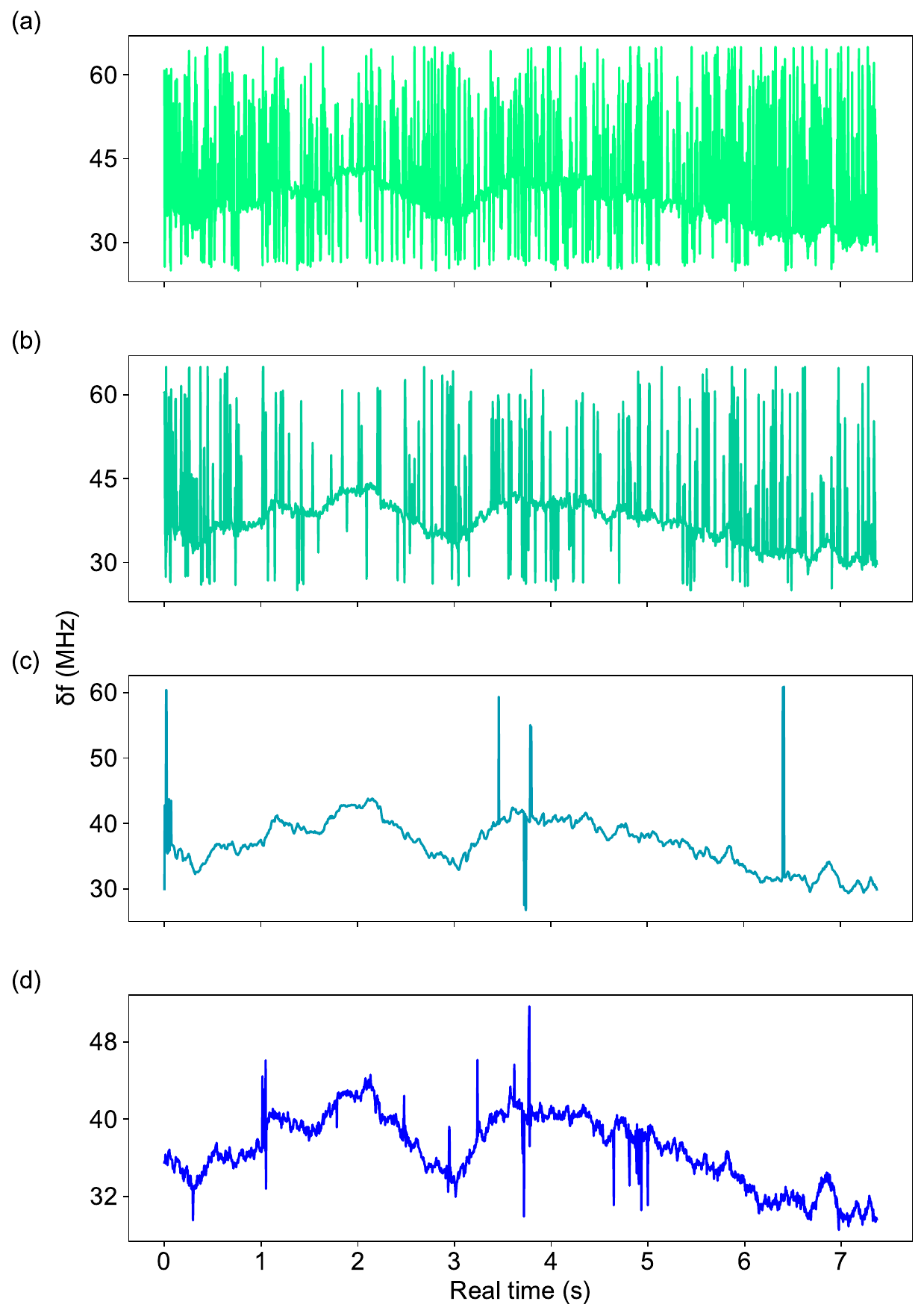}
\caption{Estimated deviations from the experimental qubit frequency, where (a), (b),
(c), and (d) are when Bayesian estimation with 1 past data, Bayesian estimation with 2
past data, and Bayesian estimation with 9 past data, the FNN with 9 past data are used
for estimation, respectively. }
\end{figure*}

\subsection{Randomized benchmarking}
In this section, we describe the protocol of RB we used. We tested two types of the Clifford composition. One is a gate set containing $\pi/2$ and $\pi$ rotations around the Bloch sphere ('standard composition (SC)' ) (Table 1(a)). There are on average 1.875 primitive gates per Clifford composition. It is widely used for RB of various kinds of qubits because it is the most efficient composition of the Clifford group in terms of gate time. A potential drawback is that there are many physical gates that make up a Clifford group, and they must be optimized independently to achieve high average control fidelity.
The other is a gate set containing $\pi/2$ rotations around the Bloch sphere ('$\pi / 2$ composition ($PC$)')(Table 1(b)). 
The gates $X_{\pi/2} \text { and } Y_{\pi/2}$ are explicitly referring to a rotation of $\pi/2$ around the x-axis and y-axis of the Bloch sphere of a single-qubit, respectively. There are on average 3.217 primitive gates per Clifford composition. Although this configuration is not optimal in terms of gate time, the small number of physical gates that make up the Clifford group allows for quick and easy gate fidelity evaluation and optimization\cite{pi2}.

Next, we describe our randomized benchmarking verification. Due to limitations in the waveform memory of our home-made AWG, observing complete decay saturation was challenging. Therefore, we chose to verify the zero-saturation behavior under relatively low-fidelity conditions. In Fig. S5(a), we performed randomized benchmarking under these conditions and confirmed that the difference between the spin-up and spin-down decay curves saturates to zero.
Fig. S5(b) compares RB results using SC and PC, with PC demonstrating higher fidelity. This difference arises because SC reflects the average gate fidelity of $\pi$, $\pi$/2, and I gates, while PC primarily reflects $\pi$/2 gate fidelity.

\begin{table}[ht]
\centering
\includegraphics[scale = 0.4]{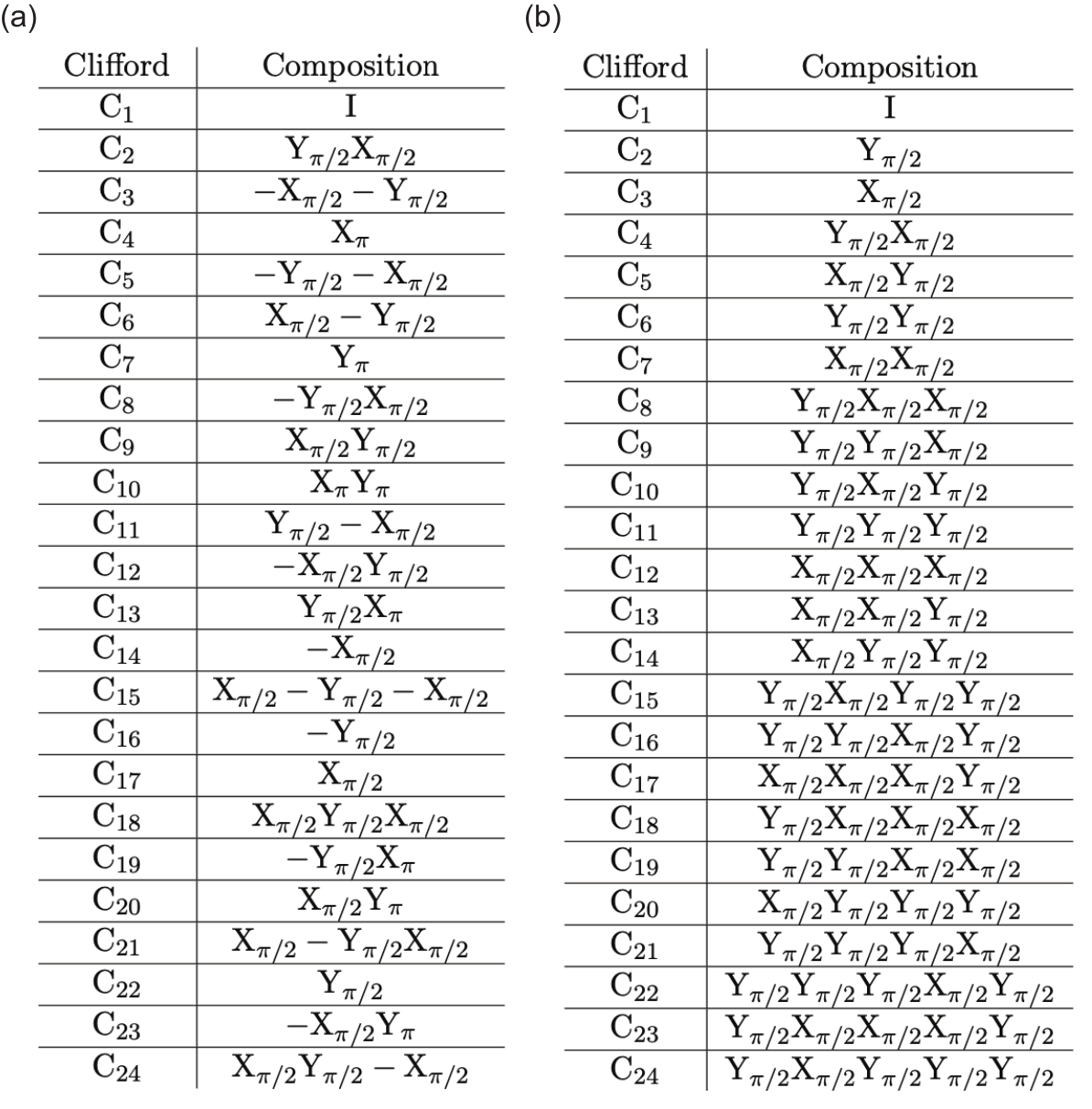}
\caption{Clifford compositions used in this work. (a) A gate set containing $\pi / 2$ and $\pi$ rotations around the Bloch sphere ('standard composition (SC)' ). (b) A gate set containing $\pi / 2$ rotations around the Bloch sphere ('$\pi / 2$ composition ($PC$)')}
\end{table}

\begin{figure*}[ht]
\centering
\includegraphics[width=0.8\textwidth]{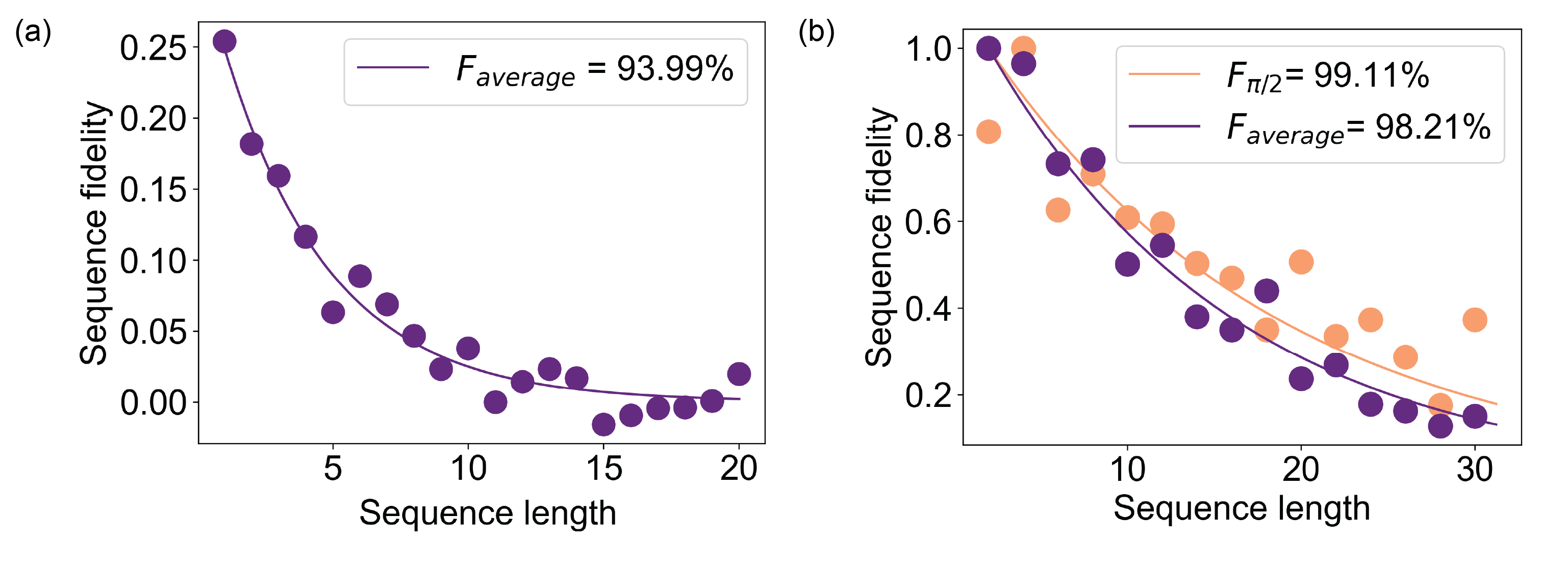}
\caption{\label{supp:fig:supp_RB} \textbf{Test runs of Randomized benchmarking} a) The sequence fidelity as a function of the number of applied clliford gates with bad fidelity condition. We confirmed that the difference between the up-spin and down-spin decay curves saturates to zero. b) RB data with different Clifford compositions (Table 1(a) and 1(b)). The difference in fidelities between the tables can be attributed to their distinct gate compositions: Table 1 provides average fidelities across all gate types including $\pi$ rotations, $\pi$/2 rotations, and identity gates, whereas Table 2 predominantly contains $\pi$/2 rotations in its compositions, thus primarily reflecting the fidelity of $\pi$/2 rotational gates.
}
\end{figure*}

\begin{figure*}[ht]
\centering
\includegraphics[width=0.6\textwidth]{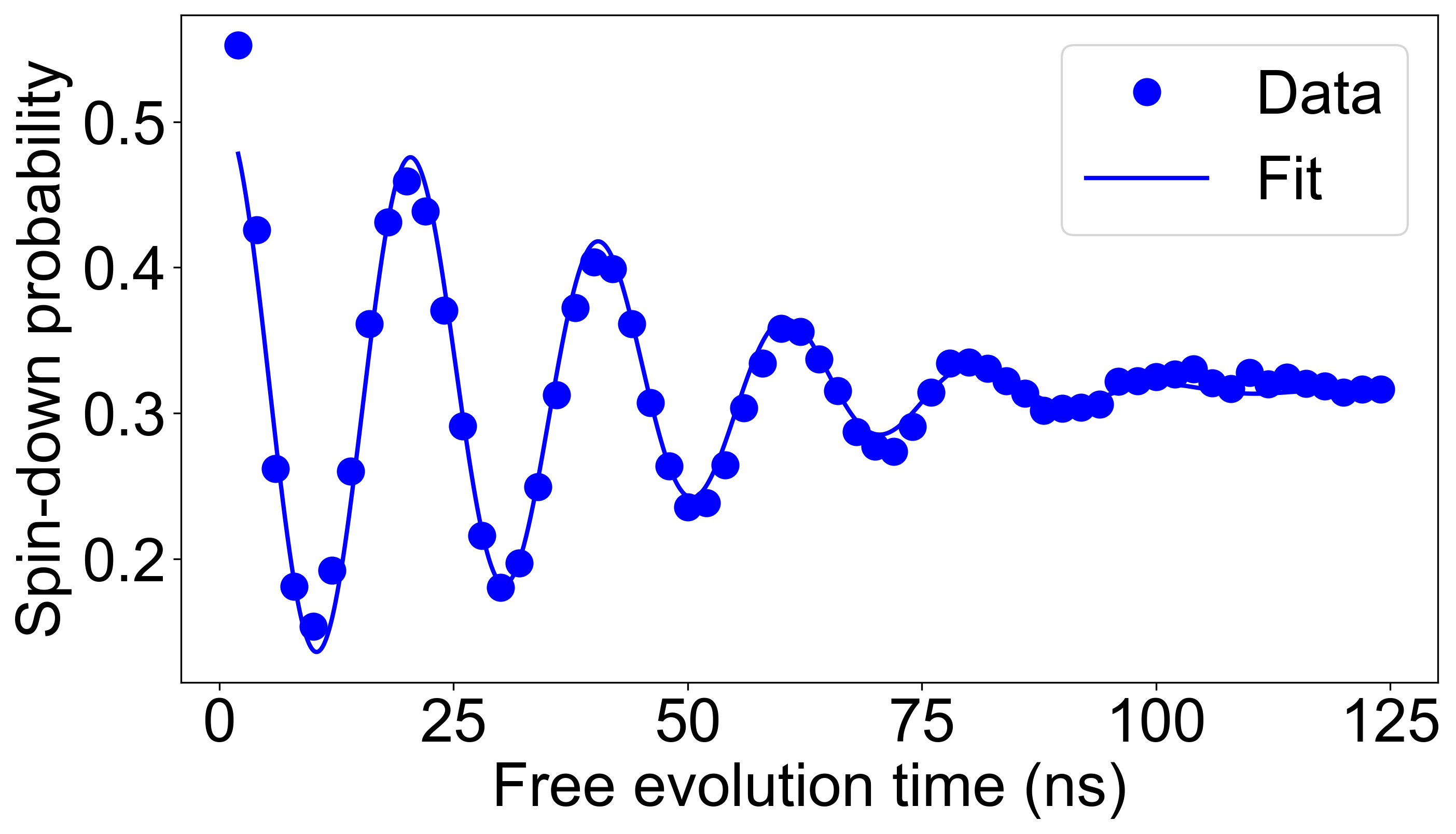}
\caption{\label{supp:fig:ramsey} \textbf{Averaged Ramsey oscillation without feedback with shorter wait time} The spin-down probability (blue circles) is plotted as a function of the free evolution time. The measurement was averaged over a same duration as that in Fig. 3(d). The solid blue line is a fit to a sinusoid with a Gaussian decay envelope, described by the function $P(t) = A \exp(-(t/T_2^*)^2) \sin(2\pi f t + \phi) + C$. From this fit, we extract a dephasing time of $T_2^* = 52.42 \pm 2.13$ ns.
}
\end{figure*}

\bibliographystyle{naturemag}
\bibliography{manualbib}
\pagebreak

\end{document}